\documentclass[aip,apl,reprint,showpacs,groupedaddress]{revtex4-1}

\usepackage{amssymb}
\usepackage{float}
\usepackage{graphicx}
\usepackage{color}
\usepackage{soul}
\usepackage[pdftex,colorlinks=true,linkcolor=red,citecolor=blue]{hyperref}

\pagecolor{white}
\begin{document}

\title[]{Boundary scattering in micro-size crystal of topological Kondo insulator SmB$_6$}

\author{Narayan Poudel$^1$}
\email{Narayan.Poudel@inl.gov}%
\author{Daniel J. Murray$^1$}
\author{Jason R. Jeffries$^2$}
\author{Krzysztof Gofryk$^1$}
\email{gofryk@inl.gov}%

\affiliation{$^1$ Idaho National Laboratory, Idaho Falls, Idaho 83415, USA}
\affiliation{$^2$ Lawrence Livermore National Laboratory, Livermore, California 94550, USA}
\date{\today}%

\begin{abstract}

We have studied the effects of phonon-boundary scattering on the thermal transport in topological Kondo insulator, SmB$_6$. The studies have been performed by using the  $3\omega$ method in the temperature range 300 K - 3 K. We show that the observed thermal conductivity of micro-size SmB$_6$ is of the order of magnitude smaller than for a bulk single-crystal. Using the Callaway model we analyzed the low-temperature lattice thermal conductivity of the micro crystal and show that phonon scattering by sample boundaries plays a major role in the thermal resistance in this topological material. In addition, we show that the temperature dependence of the lattice thermal conductivity shows a double peak structure that suggests complex phonon-phonon or phonon-defects interactions in SmB$_6$. These findings provide guidance for the understanding of the thermal transport of advanced materials and devices at a micro-scale.
\end{abstract}

\maketitle

\section{Introduction}                         

Recently, SmB$_6$ has been characterized as a new topological Kondo insulator in which a narrow bandgap opens at low temperatures due to the hybridization of conduction and valence bands, associated with the band inversion and the presence of the surface states.\cite{Menth:69, Dzero:10, Wolgast:13, Kim:14, Butch:16, Thomas:16} These findings stimulated a large volume of experimental and theoretical work on this material.\cite{ Kim:13, Zhu:13, Zhang:13, Neupane:15, Chen:15, Yasuyuki:16} In addition to  detailed electrical and magneto-transport studies, the thermal transport properties of SmB$_6$ have been evaluated to provide more insight into the low-energy excitations, magnetic field effect of surface states, and origin of possible Fermi surfaces.\cite{Boulanger:18, Hartstein:18} Furthermore, various transport studies have given conflicting results mostly due to the difficulty of obtaining good quality single crystals.\cite{Li:14, Tan:15, Neupane:15, Thomas:19} The thermal conductivity has been also studied in various topological materials to understand the heat transport phenomena of topologically protected high mobility carriers and the variation of Lorenz number, \cite{Baldomir:19, Luo:18} and simialr studies have also been performed on SmB$_6$.\cite{Flachbart:82, Popov:07, Yue:16, Boulanger:18} They have not provided, however, a decisive signature of the electronic contribution at low temperatures, most probably due to major contribution from the bulk. It has been shown recently that reducing dimensions and increasing a surface to volume ratio is a good method to probe surface states in topological systems,\cite{ Kim:11, Wang:16, Hosen:20} including SmB$_6$.\cite {Yong:14, Syers:15, Liu:18}

Various measurement techniques have been developed to accurately measure the thermal transport in a variety of materials ranging from bulk, through thin films and nano-size materials.\cite{Zhao:16, Tritt:05} For instance, the steady-state methods \cite{Pope:01, Zawilski:01} generally require large samples (to reliably measure the temperature across the crystal),  which are not always available, especially in the case of single crystals. On the other hand, transient methods \cite{Maldonado:92, Cahill:90, Parker:61} can be more convenient over the steady-state methods where the measurements are performed as a function of time. The advantage of these methods is that they can be used for both bulk materials and thin films.\cite{Zhao:16} Among several transient methods, $3\omega$ method \cite{Cahill:90} gained some special attention recently as a method of choice for thermal conductivity measurements of micro-size samples as well as thin films and even organic tissues.\cite{Lu:01, Chen:04, Choi:09, Li:13, Kim:99, Choi:13, Lubner:15} The advantage of  $3\omega$ method lies in less heat radiation during the measurements \cite{Xing:14} compared to stead state technique in particular.\cite{Shrestha:18} In this work, we used the $3\omega$ method to study the low-temperature thermal conductivity of a micro-sized single crystal of topological Kondo insulator, SmB$_6$. The results have been analyzed by the Callaway model \cite{Callaway:59} and we show that the strong phonon-boundary scattering governs the reduction of the thermal conductivity of the micro size crystal. We also observed a double peak dependence of $\kappa_{ph}(T)$ that is reminiscent of systems with resonant phonon scattering and might indicate the presence of complex phonon interactions in this topological material.

\section{Experimental}

\begin{figure}[]
\begin{center}
\includegraphics[angle=0, width=3.3 in]{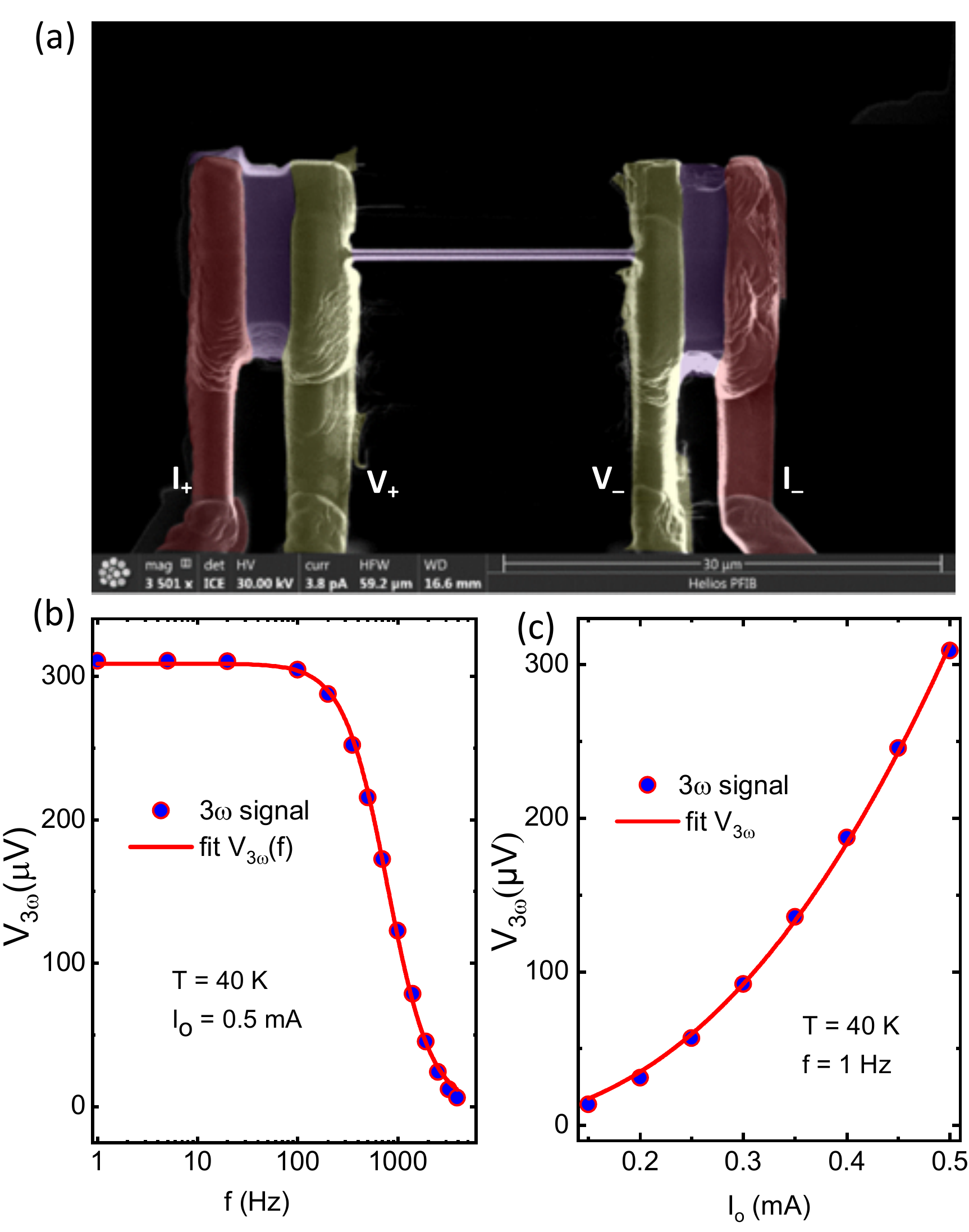}
\end{center}
\caption{(a) Image of pFIB prepared SmB$_6$ lamella with four electrical contacts (platinum) for $3\omega$ measurements. The outer two contacts are current leads the and inner two contacts are voltage leads (see text). The dimensions of the sample are 20 $\times$ 0.6  $\times$ 7 $\mu m^3$. (b) Variation of  $3\omega$ voltage ($V_{3\omega}$) versus frequency of SmB$_6$  at 40 K and at  constant electrical current of 0.5 mA. The solid line is the least-square fit of  Eq.\ref{eq1}. (c) Variation of  $3\omega$ voltage versus amplitude of AC current taken at 40 K while the frequency of the current is fixed at 1 Hz. The solid line is the least-square fit to $V_{3\omega}  \propto I_o^{n}$ (see text).}
\label{fig1}
\end{figure}

Polycrystalline samples of SmB$_6$ were prepared by arc melting of stoichiometric amounts of samarium and boron. More details on the sample preparation and characterization can be found in Ref. \onlinecite{Poudel:20} (the same samples have been used in the present studies). Small grains of SmB$_6$ sample were identified and micro-machined by shielded FEI Helios plasma focused ion beam (pFIB). Four platinum contacts were deposited on the SmB$_6$ lamella for standard four-probe transport measurements with the heat current along $\textless$100$\textgreater$ direction as shown in Fig.\ref{fig1}(a). Extensive imaging was avoided to prevent damaging of the sample.\cite{Zhou:05} A Quantum Design DynaCool-9 system has been used to control the sample temperature in a high vacuum condition while measuring the $3\omega$ voltage. The current value was kept low ($\textless$1 mA) during the measurements to avoid heating effects as well as to protect the sample contacts. In the current studies, we have used an experimental configuration described in more detail in Ref. \onlinecite{Shrestha:18}. 

\section{Results and Discussions}

\begin{figure}
\begin{center}
\includegraphics[angle=0, width=5.05 in]{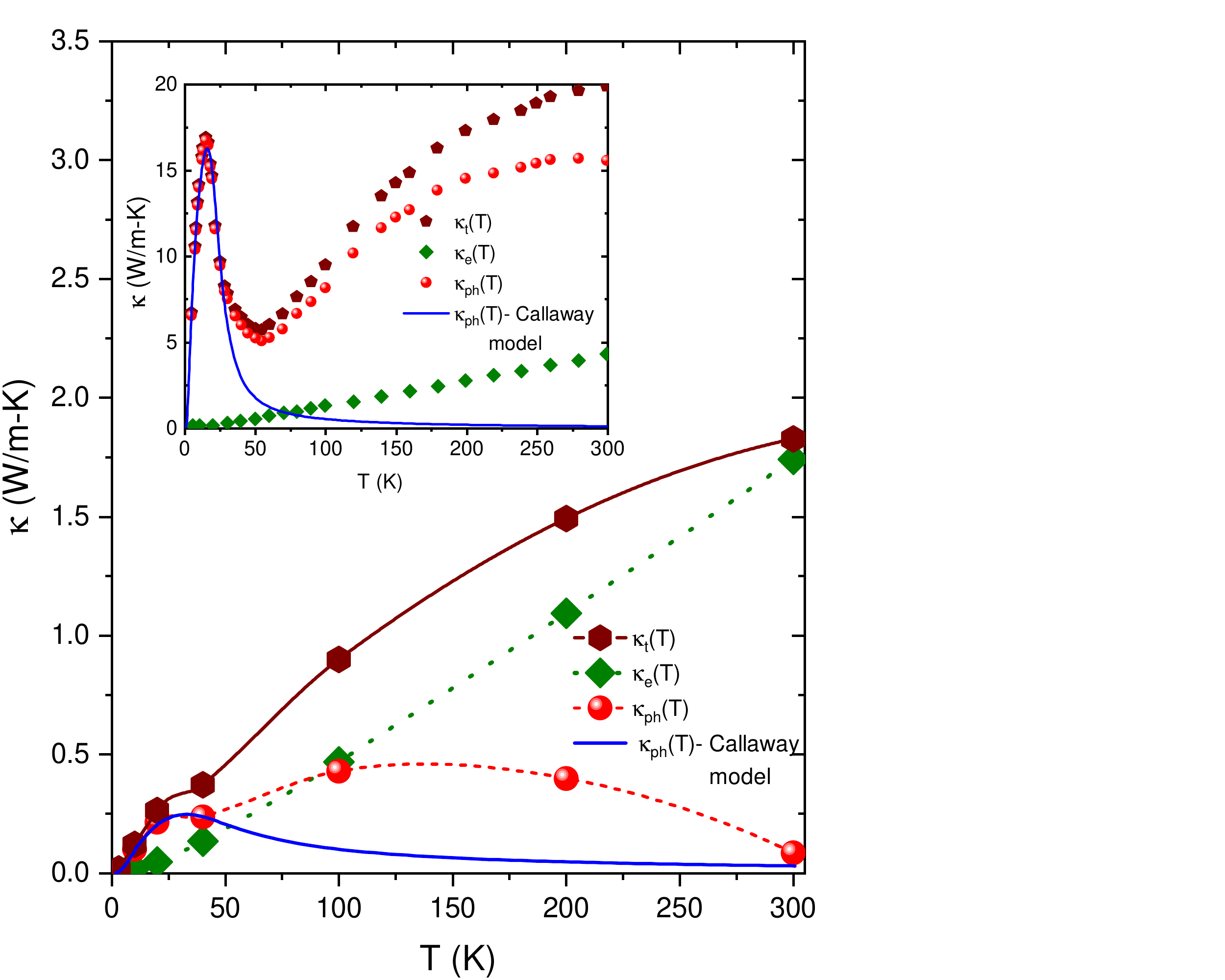}
\end{center}
\caption{Temperature dependance of thermal conductivity of SmB$_6$ lamella. The $\kappa_{e}$ and $\kappa_{ph}$ represent the electronic and phonon contributions (see text). The solid blue line is an analysis of $\kappa_{ph}$ by the Callaway model below 40 K and extrapolated to 300 K. The inset shows the thermal conductivity data for bulk SmB$_6$ single crystal. The data has been taken from Ref. \onlinecite{Popov:07}. Similarly, the solid blue line represents the Callaway model fit to $\kappa_{ph}$ below 40 K and extrapolated to 300 K.}
\label{fig2}
\end{figure}

In general, when the AC current of a certain frequency is applied across the sample (see for instance the sample in Fig.\ref{fig1}(a)), the voltage with the same frequency ($1\omega$), as well as higher harmonics signals, are generated. The $3\omega$ voltage is generally about three orders of magnitude smaller than $1\omega$ voltage and several methods have been used to overcome this issue. Considering a one-dimensional heat flow, the $3\omega$ voltage can be written as: \cite{Lu:01}

\begin{equation}
V_{3\omega}  \approx \frac{\sqrt2I_o^3RR^{'} L}{\pi^4 \kappa S \sqrt{1+(2\gamma\omega)^2}}, 
\label{eq1}
\end{equation}

where $I_o$  is the amplitude of the AC current, R is the resistance of the sample, $R^{'} = dR/dT$ is the derivative of resistance with temperature, $L$ is the length of the sample, $S$ is the cross-sectional area of the sample, $\omega = 2\pi f$ is the angular frequency of AC current, $\gamma$ is the thermal time constant and $\kappa$ is the thermal conductivity of the sample. As seen from the expression; $V_{3\omega} \propto \frac{1}{\sqrt{1+(2\gamma\omega)^2}}$ and $V_{3\omega} \propto I_o^3$. Therefore, the $3\omega$ voltage can be obtained either by varying the frequency of the current or by varying the amplitude of the current. As a typical example, the variation of the $3\omega$ voltage with frequency for a fixed current of $I_o $=0.5 mA at T = 40 K of the SmB$_6$ sample is shown in Fig.\ref{fig1}(b). The  $V_{3\omega}$ signal decreases slightly at lower values of $f$ and decreases rapidly with increasing $f$ as displayed in the semilogarithmic plot. The solid line is the least-square fit to the data using Eq. \ref{eq1}, where the $\kappa$ and $\gamma$ are treated as free parameters. A similar approach has been performed for each temperature to obtain the thermal conductivity of SmB$_6$ microcrystal (see Fig.\ref{fig2}). The variation of the $V_{3\omega}$ with $I_o$ at low frequency ($f$ = 1 Hz)  is shown in Fig.\ref{fig1}(c). It is found that $V_{3\omega}$ varies with $I_o^{2.4}$, which is close to the expected cubic dependence. For low frequency ($\gamma\omega \rightarrow 0$), the  $V_{3\omega}$ voltage can be fitted with $V_{3\omega}  \approx \frac{4 I_o^3RR^{'} L}{\pi^4 \kappa S}$ to get $\kappa$.\cite{Lu:01} By the least-square fit to the experimental data, treating $\kappa$ as the free parameter, we have obtained similar values of $\kappa$ to the ones obtained by varying the frequency. This gives us additional confidence of the measured thermal conductivity of our sample.

The temperature dependence of the thermal conductivity of SmB$_6$ is shown in Fig.\ref{fig2}. The highest value of $\kappa_t$ is observed at 300 K, which is 1.83 W/m-K and it decreases when the temperature is lowered. In general, in non-magnetic systems the $\kappa(T)$ contains two major contributions; the electronic contribution ($\kappa_e$) and phonon contribution ($\kappa_{ph}$), so that $\kappa{_t}(T)$ = $\kappa_{e}(T)$ + $\kappa_{ph}(T)$. The electronic contribution can be estimated using the Wiedemann-Franz law; $\kappa_{e}(T) = 2.44\times10^{-8}$ $T/\rho(T)$,  \cite{Franz:53} where $T$ stands for temperature and $\rho(T)$ is the electrical resistivity of the sample (Ref. \onlinecite{Poudel:20}). By subtracting the $\kappa_e$ from $\kappa{_t}(T)$, the $\kappa_{ph}$ can be obtained, which is also shown in Fig.\ref{fig2}. Interestingly, at room temperature the so-obtained  $\kappa_e$ is slightly lower than $\kappa(T)$, indicating that the majority of heat is carried by electrical carries. The thermal conductivity measurements obtained on a micro-size sample differ from the one obtained on a mm size single crystal. For comparison we have included results of SmB$_6$ single crystal that is shown in the inset of Fig.\ref{fig2} (data taken from Ref. \onlinecite{Popov:07}). As can be seen from the figure the thermal conductivity of the single crystal is an order of magnitude larger than obtained for micro size crystal. At low temperatures (below 50 K) there is a characteristic anomaly in $\kappa_{ph}(T)$ that results from defect, boundary, phonon-phonon scattering. In general, in non-magnetic insulators, phonon-phonon interactions lead to a reduction of the thermal conductivity with an increase of temperature.\cite{Tritt:05} In SmB$_6$, however, the thermal conductivity increases with increasing temperature, creating a maximum at around 150 and 300 K respectively for micro and bulk crystals. The double peak temperature dependence of the thermal conductivity of SmB$_6$ is atypical for insulating, non magnetic crystals, where a single peak in $\kappa (T)$ is observed at low temperatures.\cite{Tritt:05} Interestingly,  similar behavior of $\kappa (T)$ has been observed in the systems where resonant phonon scattering is present in the crystal.\cite{Phol:62, Verma:72, Gofryk:14, Liu:16} If so, this might indicate a presence of complex vibrational properties in SmB$_6$ and its strong coupling to other degrees of freedom such as potential magnetic and non-magnetic defects or low energy phonon branches.\cite{Phol:62, Verma:72, Gofryk:14, Liu:16}

In order to analyze the lattice thermal transport of SmB$_6$ at low temperatures, we use the Callaway model.\cite{Callaway:59} In this approach the phonon thermal conductivity can be expressed as: \cite{Callaway:59}

\begin{equation}
\kappa_{ph}(T) = \frac{k_B^4}{2\pi^2\hbar^{3}v_{ph}}T^3\int_{0}^{\frac{\Theta_D}{T}} \frac{x^4e^x}{(e^x-1)^2}\tau(\omega, T) dx,
\end{equation}

where $x=\frac{\hbar\omega}{k_BT}$, $v_{ph}$ is the average phonon velocity, and $\Theta_D$ is the Debye temperature. The frequency-dependent relaxation time of phonons $\tau(\omega, T)$ contains several terms of scattering mechanism. For our analysis we considered  $\tau(\omega, T) = [\frac{v_{ph}}{L_b}+D\omega^4+U\omega^2 T \mathrm{ exp}\left(-\frac{\Theta_D}{3T}\right)]^{-1}$ where  $B=\frac{v_{ph}}{L_b}$ is the boundary scattering term with ${L_b}$ that represents the smallest dimension of sample, $D\omega^4$ is the defect scattering term, and $U\omega^2 T \mathrm{ exp}\left(-\frac{\Theta_D}{3T}\right)$ is the Umklapp scattering term. Using phonon velocity $v_{ph}$ = 5000 m/s, \cite{Boulanger:18,Nakamura:91} and Debye temperature $\Theta_D$ = 373 K,\cite{smith:85} we have analyzed the experimental low temperature $ \kappa_{ph}(T)$ dependence of bulk and micro size SmB$_6$, as shown in Fig.\ref{fig2}. From the analysis, we found that the U parameter responsible for the phonon-phonon Umklapp scattering is similar for the two samples and equals to $0.44\times10^7$ $K^{-3}S^{-1}$ and $1.27\times10^7$ $K^{-3}S^{-1}$, for bulk and micro single crystals, respectively. This is somehow expected at this temperature range since the U processes dominate the thermal resistance at high temperatures. The largest differences are in defect and especially boundary scattering terms. The so-obtained $D$ and $L_b$ parameters are: 1100  $K^{-4}S^{-1}$ and 0.171 $mm$ for bulk and 26242 $K^{-4}S^{-1}$ and 0.211 $\mu m$ for micro samples. The $L_b$ parameter is an estimation of the smallest dimension of the sample for single crystals materials. As can be seen, the obtained parameters are of the same order as the smallest dimensions of the measured samples ($\sim$0.1 $\mu m$ in case of our FIB sample and $\sim$1 $mm$ for the single crystal \cite{Popov:07}). This indicates that the boundary scattering is prevalent in our sample and plays a significant role in reducing the  $ \kappa_{ph}$, as compared to the bulk single crystal. 

\begin{figure}[h]
\begin{center}
\includegraphics[angle=0, width=3.3 in]{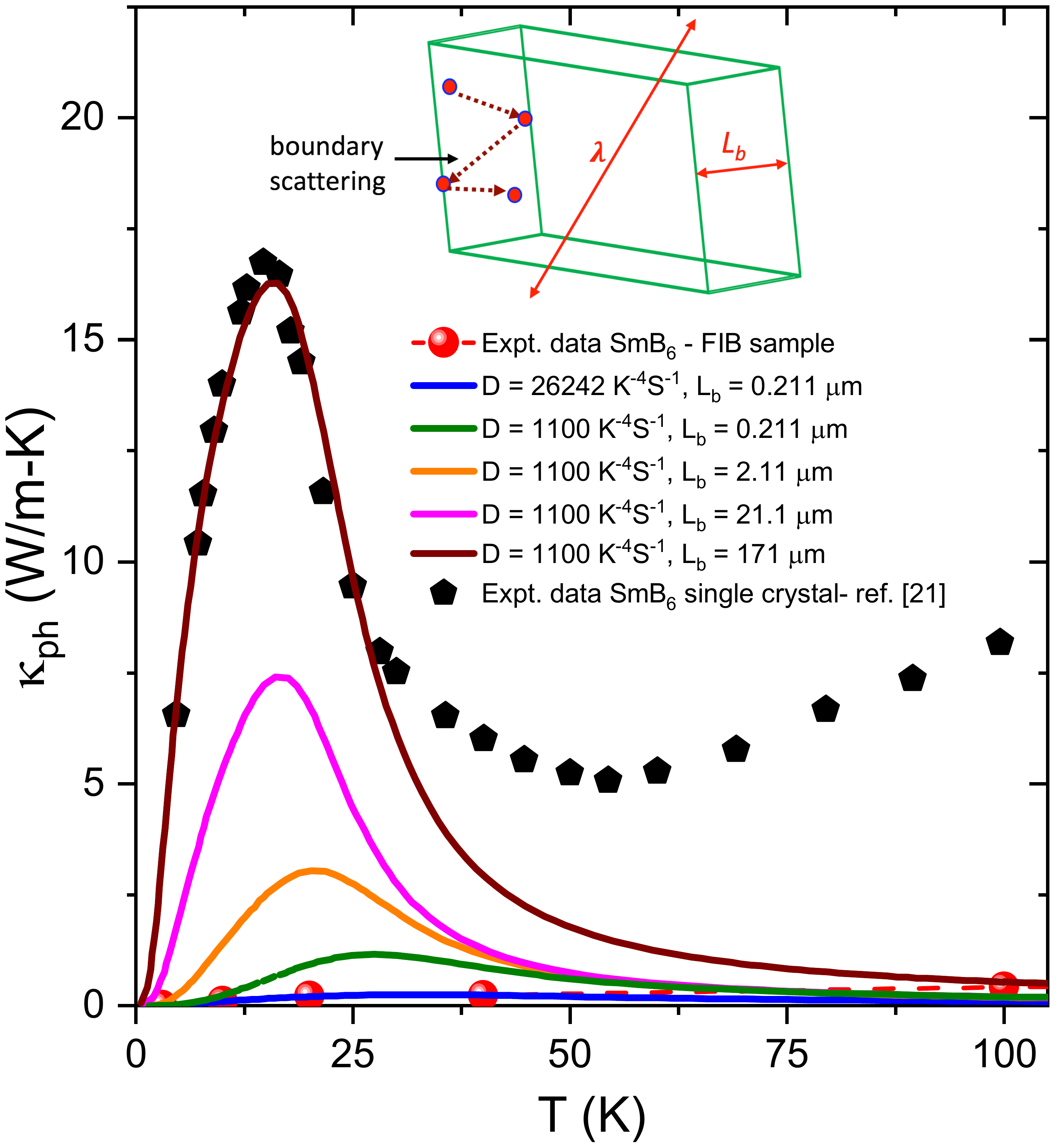}
\end{center}
\caption{(Color online) The low-temperature dependence of the phonon part of the total thermal conductivity of SmB$_6$. The data for the SmB$_6$ lamella is represented by solid symbols and the solid squares are the data for bulk single crystals (data taken from Ref. \onlinecite{Popov:07}). Solid lines represent the modeled curves using the Callaway model for different parameters $D$ and $B$ (see text). The inset shows the sketch of a phonon scattering from the sample boundary.}
\label{fig3}
\end{figure}

Our analysis points to a slightly larger effect of defect scattering in our micro-size crystal than in the bulk single crystal. This may be directly related to different synthesis methods used and different amounts of defects centers present in the samples. It has been shown that SmB$_6$ system is very sensitive to synthesis methods that result in a different quality of the samples.\cite{Li:14, Tan:15, Thomas:19} To look at this phenomenon in more detail we have modeled the thermal conductivity of our sample using the same value of $D$ parameter as found for the single crystal material. As can be seen in the Fig.\ref{fig3} by doing that, the thermal conductivity slightly increases as expected, but the so-obtained $\kappa_{ph}$ value at the peak reaches only 1.15~W/m-K, far away from $\sim$16 W/m-K observed for the single crystal sample (see the green curve in Fig.\ref{fig3}). This indicates that the majority of the heat reduction observed for the micro size crystals comes from the boundary scattering. When the temperature lowers, the phonon mean free path increases and may become comparable to the sample dimension and is one of the reasons that prevent phonon thermal conductivity to increase at low temperatures.\cite{Tritt:05} This behavior is observed in various nano wires, thin films and polycrystalline samples when the boundary scattering dominates the thermal transport.\cite{Li:03,  Zhang:06, Cheatio:12, Li:13, Cahil:03, Shrestha:19} In order to test this behavior, in Fig.\ref{fig3}, we show a modeled curves for the lattice thermal conductivity of SmB$_6$ by taking into account different characteristic sample sizes. During the process we kept the defect scattering term fixed. As can be seen, the low-temperature lattice thermal conductivity of SmB$_6$ microcrystal can be well described by this approach, and indicates that majority of the thermal resistance at low temperatures comes from the boundary scattering. \\

\section{Conclusion}

In summary, we have synthesized and micro-machined a $\mu$-size crystal of SmB$_6$. Using the $3\omega$ technique we have studied the effect of various scattering mechanisms and their impact on the thermal conductivity in the micro-size sample. The obtained thermal conductivity is found to be reduced by an order of magnitude than the previously reported data on bulk single crystals. We analyzed the low-temperature phonon thermal conductivity of our sample using the Callaway model and we show that the phonon boundary scattering plays the main role in reducing the thermal conductivity of this material. The temperature dependence of $\kappa_{ph}(T)$ points to complicated phonon interactions in this topological material that are characteristic of resonant scattering.  Our work also demonstrates the successful integration of the $3\omega$ method that is used to study the thermal properties of micro-size samples. This approach can be used for materials where large single crystals are not available or growing of single crystals is impossible. Also, it will play an important role in understanding the thermal conductivity of micro-size materials, where phonon boundary scattering dominates thermal resistance, especially in advanced materials and technologies such as microelectronics, thermoelectric, and nuclear research.

\section*{Data availability}
The data that support the findings of this study are available from the corresponding author upon reasonable request.

\begin{acknowledgments}
This work was supported by  Idaho National Laboratory's Laboratory Directed Research and Development (LDRD) program (18P37-008FP). D. Murray acknowledges the financial support from U.S. Department of Energy, Office of Nuclear Energy under DOE Idaho Operations Office Contract DEAC07-051D14517 as part of a Nuclear Science User Facilities.Portions of this work were performed under the auspices of the U.S. Department of Energy by Lawrence Livermore National Laboratory under Contract DE-AC52-07NA27344.
\end{acknowledgments}

\end{document}